\documentclass[12pt,preprint]{aastex}

\begin{document}

\title{Metallicities of M Dwarf Planet Hosts from Spectral Synthesis}

\author{Jacob L. Bean}
\affil{Dept.\ of Astronomy, University of Texas, 1 University Station, C1402, Austin, TX 78712}
\email{bean@astro.as.utexas.edu}

\author{G. Fritz Benedict, Michael Endl}
\affil{McDonald Observatory, University of Texas, 1 University Station, C1402, Austin, TX 78712}
\email{fritz@astro.as.utexas.edu, mike@astro.as.utexas.edu}

\received{}
\revised{}
\accepted{}

\begin{abstract}
We present the first spectroscopic metallicities of three M dwarfs with known or candidate planetary mass companions. We have analyzed high resolution, high signal-to-noise spectra of these stars which we obtained at McDonald Observatory. Our analysis technique is based on spectral synthesis of atomic and molecular features using recently revised cool-star model atmospheres and spectrum synthesis code. The technique has been shown to yield results consistent with the analyses of solar-type stars and allows measurements of M dwarf [M/H] values to 0.12 dex precision. From our analysis, we find [M/H] = -0.12, -0.32, and -0.33 for GJ 876, GJ 436, and GJ 581 respectively. These three M dwarf planet hosts have sub-solar metallicities, a surprising departure from the trend observed in FGK-type stars. This study is the first part of our ongoing work to determine the metallicities of the M dwarfs included in the McDonald Observatory planet search program.

\end{abstract}


\section{INTRODUCTION}

The detection of the first candidate planetary mass companion orbiting a solar-type star \citep{mayor95} ushered in the era of extrasolar planet research in astronomy. Tremendous progress in this area has been made in the decade since and more than 200 candidate planets have been announced. Despite the success of the current detection methods, unambiguous direct imaging of an extrasolar planet orbiting a star is still an elusive goal. Therefore, most of the knowledge regarding planet formation and evolution that has been garnered from the systems detected so far has come from statistical studies of the system properties and the host stars themselves. One of the interesting findings from these studies is the trend toward higher photospheric metal abundances in extrasolar planet hosts stars relative to stars without detected planets.

\citet{gonzalez97} first noted the high metallicities of the first four stars which were found to exhibit radial velocity  variations attributable to a planetary mass companion. This trend was found to continue as more stars were identified as potential extrasolar planet hosts with the Doppler method, and followed up with high precision abundance analyses \citep{fuhrmann97, fuhrmann98, gonzalez98, gonzalez99, gonzalez01, santos00, santos03, santos04, santos05, laws03, fv05}. The most likely explanation for this observed trend is the so-called ``primordial'' hypothesis. That is, the high photospheric metal abundances in the host stars are relics of protostellar clouds and disks with a proportionally high metal content \citep{santos04, fv05}. It is theorized that high-mass planet formation is increased in high metal-content protoplanetary disks under the core-accretion paradigm \citep{pollack96}. This hypothesis explains why more Jupiter and higher-mass planets have been detected around stars with high metallicities. 

The majority of extrasolar planets have been found around FGK-type stars as these are the stellar types that make up the majority of targets in Doppler surveys. High precision abundance analyses for these types of stars are relatively straightforward and, therefore, these are the types of stars for which the metallicity -- giant planet connection has been established. However, the majority of stars in the solar neighborhood are M dwarfs \citep{henry98} and a complete understanding of planet formation must necessarily include late-type stars. 

To date, only one M dwarf, GJ 876, is known to harbor a Jupiter-mass companion \citep{delfosse98,marcy98,benedict02}. In addition to the astrometrically confirmed outer planet, another Jupiter-mass planet and a very low-mass planet in shorter-period orbits have been detected around GJ 876 \citep{marcy01, rivera05} making it also the only M dwarf multi-planet host known to date. The M dwarfs GJ 436 \citep{butler04} and GJ 581 \citep{bonfils05b} are hosts to candidate Neptune-mass planets in short-period orbits based on their radial velocity variations. Other planets, including a gas giant planet and a $\sim$ 5.5 $\mathcal{M}_{\earth}$ planet, have been detected around suspected M dwarfs using the microlensing technique \citep{beaulieu06,gould06}. These host stars are still confused with the source stars and unavailable for further study. The first star that was identified as an extrasolar planet host with the microlensing technique was thought to be an M dwarf \citep{bond04}, but recent observations have shown that it is actually a K dwarf \citep{bennett06}.

In this Letter, we present the results of our abundance analysis of three M dwarfs with a known (GJ 876) or candidate (GJ 436 and GJ 581) planetary mass companion. In \S 2 we describe our high resolution (R) and signal-to-noise (S/N) spectroscopic observations. We discuss our analysis and present our results in \S 3. We briefly discuss our results and ongoing effort to determine the metallicities for all the M dwarfs being monitored in a Doppler survey in \S 4.

\section{OBSERVATIONS AND DATA REDUCTION}
We observed GJ 876 and 436 using the 2.7 m Harlan J. Smith telescope at McDonald Observatory on November 20, 2003 and January 24, 2005 respectively. Data were obtained with the 2dcoud\'{e} spectrograph \citep{tull95} equipped with a 79 gr mm$^{-1}$ echelle grating and 8.2\arcsec x 1.2\arcsec\ slit. Two 30 minute exposures were taken for both objects and co-added before order extraction. 

GJ 581 was observed using the 9.2 m effective aperture Hobby-Eberly Telescope (HET) at McDonald Observatory on May 11, 2006 with the High Resolution Spectrograph \citep[HRS][]{tull98} fed by a 2\arcsec optical fiber. The HRS was used in the ``R = 60,000'' mode with a 316 gr mm$^{-1}$ cross-dispersion grating. The cross-dispersion grating was positioned so that the break between the two CCD chips was at 7940 \AA. Two 10 minute exposures were taken for GJ 581 and co-added before order extraction.

CCD reduction and optimal order extraction were carried out using the REDUCE package \citep{piskunov02}. The wavelength calibrations for each object were calculated based on the identification of roughly 1000 lines in thorium-argon emission spectra taken at the beginning of each respective night and have RMS precisions of 0.002 \AA. The final one-dimensional spectra of GJ 876, 436, and 581 have S/N, of 430, 360, and 190 pixel$^{-1}$ respectively at 8700 \AA. The measured resolving powers were roughly 50,000 for the 2.7m and 60,000 for the HET data.

\section{ANALYSIS AND RESULTS}
We analyzed the observed spectra of the M dwarf planet hosts using the technique described by \citet{bean06} to determine their metallicities. \citet{bean06} utilized analyses of both components of solar-type and M dwarf visual binaries to test and improve their spectroscopic analysis technique and cool-star model atmospheres. Their test was based on the assumption that unevolved stars in bound systems have the same photospheric abundances. \citet{bean06} showed that their method yields metallicities for M dwarfs consistent with the results given by standard analysis techniques applied to solar-type stars. We briefly describe the technique here, and refer the reader to that paper for a complete description.

Our analysis of the M dwarf planet hosts relied on fitting synthetic spectra to their observed spectra. We used spectral regions containing a strong TiO bandhead and relatively clean atomic line profiles as the constraints and the $\chi^{2}$ statistic as our goodness of fit metric. We generated synthetic spectra with an updated version of the plane-parallel, local thermodynamic equilibrium (LTE), stellar analysis code MOOG \citep{sneden73}. We adopted a grid of model atmospheres which were computed for this particular purpose using the model atmosphere code PHOENIX \citep[version 13,][]{haus99}. We generated a model atmosphere with arbitrary parameters to be input into MOOG by interpolating in this grid. 

We determined the stellar parameters effective temperature, {T}$_{eff}$, metallicity, [M/H]\footnote{We adopt the standard spectroscopic notation: for elements X and Y, log $\epsilon$(X) $\equiv$ log$_{10}(N_{X}/N_{H})$ + 12.0, [X/Y] $\equiv$ log$_{10}(N_{X}/N_{Y})_{\star}$ - log$_{10}(N_{X}/N_{Y})_{\sun}$, and $N_{X}$ is the number density of element X.}, and microturbulent, $\xi$, and macroturbulent, $\eta$, velocities directly from the spectral fitting. We constrained the stellar surface gravity using an empirical log \textsl{g} -- $\mathcal{M}$ relationship that was derived from recent measurements of M dwarf radii. We estimated the masses of the objects to use with this relationship based on the M$_{K}$ -- $\mathcal{M}$ relationship in \citet{delfosse00} and using the \textsl{K} magnitudes and parallaxes from the Two Micron All Sky Survey (2MASS) point source catalog \citep{cutri03} and the Hipparcos catalog \citep{hipp, perryman97} respectively. We assumed the abundances given by \citet{asplund05} as the reference solar abundances and linear alpha element (O, Mg, Si, Ca, and Ti) and C enhancement relationships as functions of [Fe/H] from the data presented by \citet{carlos04}. 

The results from our analysis of the M dwarf planet hosting stars GJ 876, 436, and 581 are given in Table ~\ref{tab:table1}. We find [M/H] = -0.12, -0.32, and -0.33 for GJ 876, GJ 436, and GJ 581 respectively. We adopt the standard uncertainties in the parameters derived using this technique as given by \citet{bean06}. They are 48 K, 0.10 dex, 0.12 dex, 0.15 km s$^{-1}$, and 0.20 km s$^{-1}$ for {T}$_{eff}$, log \textsl{g}, [M/H], $\xi$, and $\eta$ respectively. These uncertainties, which we consider to be inherent in the analysis technique, were calculated based on the agreement between the M dwarf and solar-type star visual binary pairs in the \citet{bean06} analysis. This external error estimation method includes correlated uncertainties from all the other determined parameters and yields uncertainties that are more conservative and robust than errors derived from only considering the uncertainty in the fitting process from the spectral S/N.

\begin{figure}
\epsscale{1.0}
\plotone{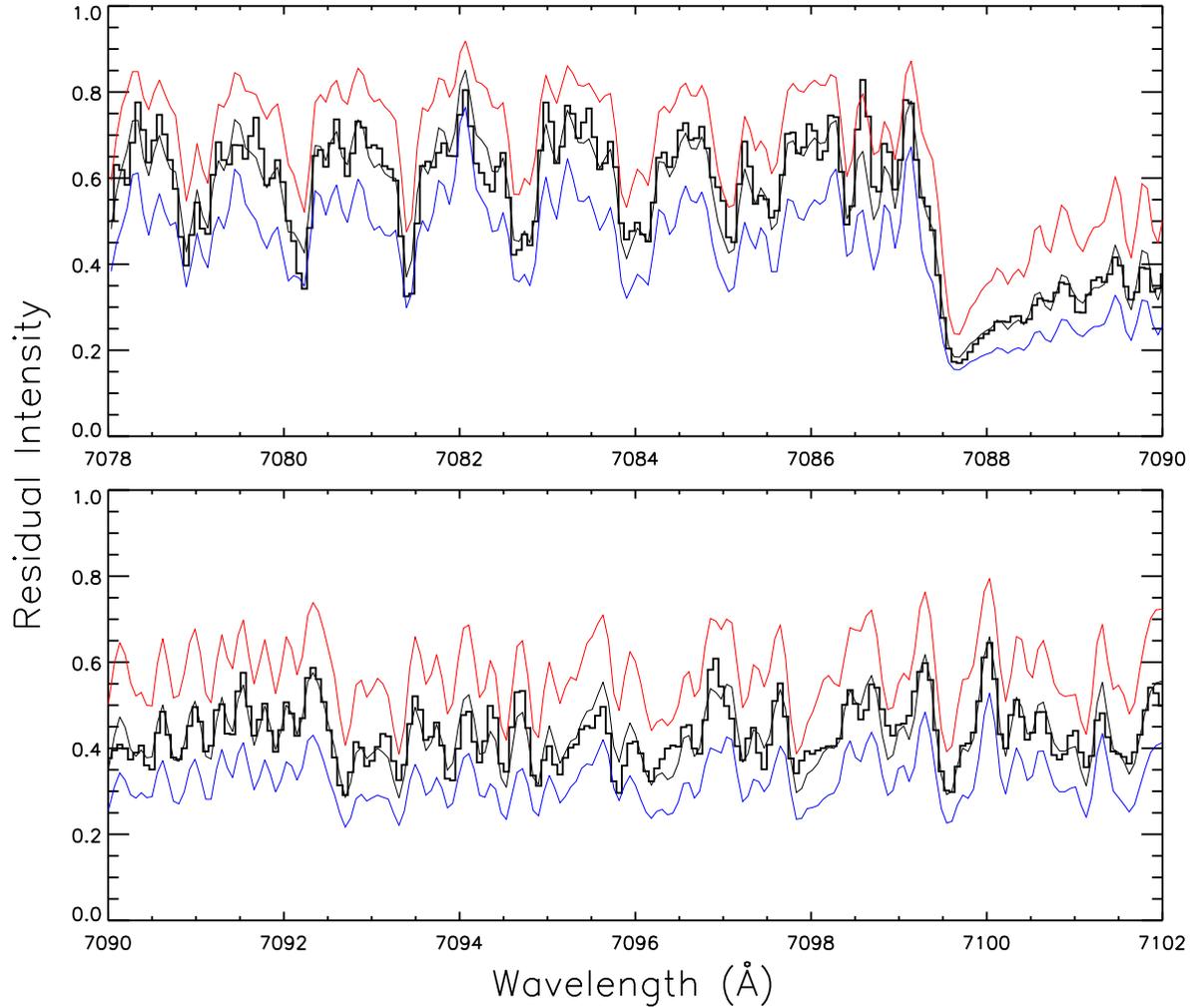}
\caption{Spectral region near the strong TiO $\gamma$ R$_{2}$ 0 -- 0 bandhead for GJ 876 (histogram). The best fit used to determine the stellar parameters is over-plotted (solid black line). For comparison, synthetic spectra computed with [M/H] values 0.3 dex lower (dotted line, solid red in the electronic edition) and higher (dashed line, solid blue in the electronic edition) than the best fit value are also over-plotted.}
\label{fig:f1}
\end{figure}

\begin{figure}
\epsscale{1.0}
\plotone{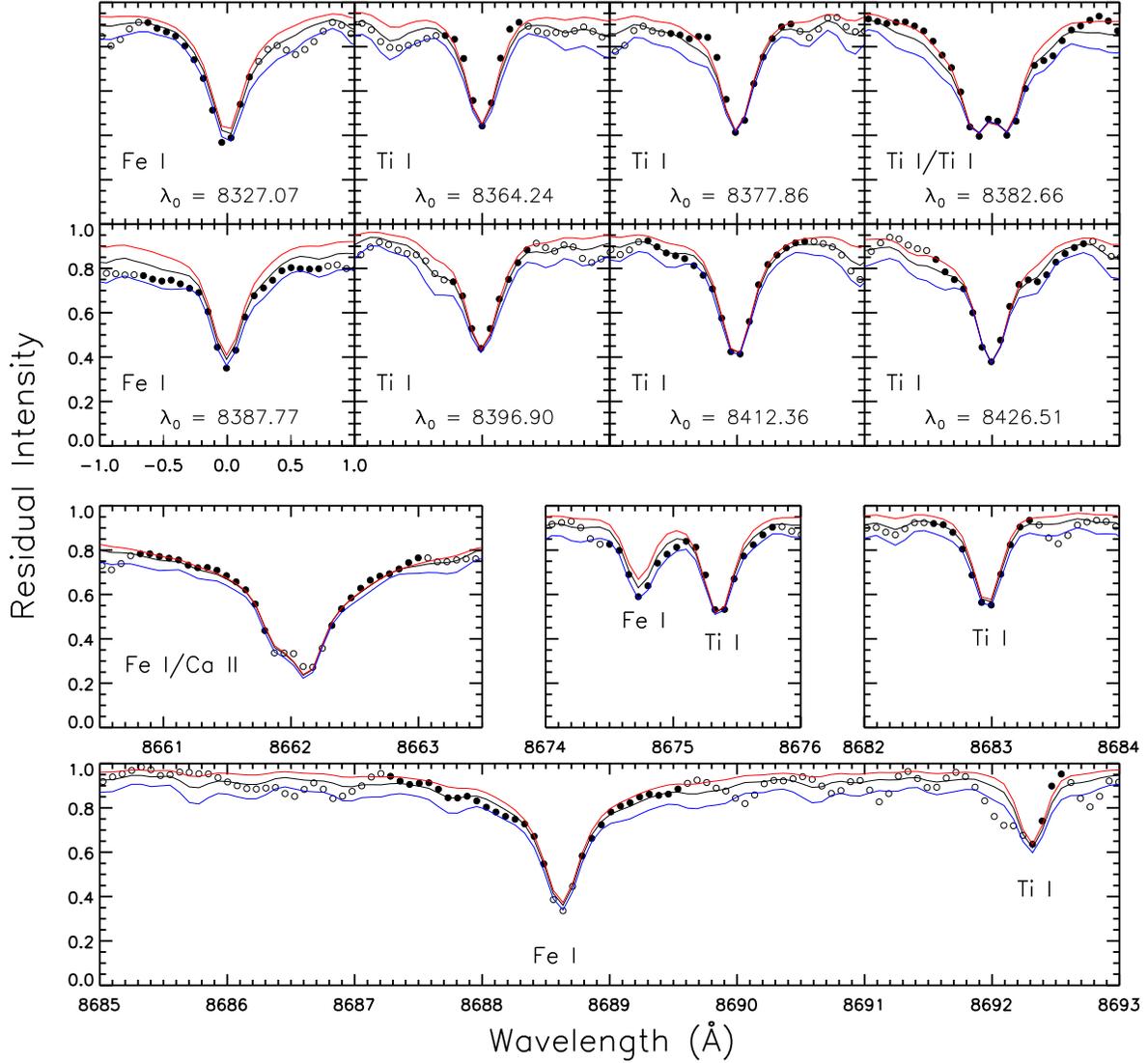}
\caption{Fit of synthetic spectra (solid line) to atomic line profiles (points) for GJ 876. The filled points were used in the fitting process; the open points were ignored. For comparison, synthetic spectra computed with [M/H] values 0.3 dex lower (dotted line, solid red in the electronic edition) and higher (dashed line, solid blue in the electronic edition) than the best fit value are also over-plotted. The panels are sorted by wavelength and the linear scaling in both parameters is the same throughout. The lines in each half, top and bottom, make up a contiguous spectral order in our observed spectra. All apparent ``lines'' in the figure that aren't fit are actually multiple TiO lines.}
\label{fig:f2}
\end{figure}

Plots of the observed spectrum and best fit synthetic spectrum for GJ 876 are shown in Figures ~\ref{fig:f1} and ~\ref{fig:f2}. Synthetic spectra computed with [M/H] values 0.3 dex lower and higher than the best fit value are also included in these Figures to illustrate the sensitivity of our measurement technique. The ``high'' and ``low'' metallicity synthetic spectra clearly do not match the observed spectrum as well as the synthetic spectrum computed with the determined stellar parameters.

An interesting aspect of this result is the closeness of the derived {T}$_{eff}$ values for all three stars (range of 20 K) despite a range of 1.5 spectral types. The explanation for this is that metallicity and effective temperature are degenerate in the M dwarf spectral classification system. Therefore, only a detailed analysis such as the one we have employed can break the degeneracy and give a precise estimate of these parameters for an M dwarf.

\section{DISCUSSION}
\citet{butler04} report an occurrence rate of Jupiter-mass planets (0.5 $\mathcal{M}_{Jup} < \mathcal{M} < $13 $\mathcal{M}_{Jup}$) with orbital semi-major axes, $a$, $<$ 1 AU of 3.5\% around FGK-type stars. In contrast, \citet{endl06} found a frequency of 0.46\% with an upper limit of 1.27\% for the occurrence of Jupiter-mass planets around M dwarfs based on a dedicated survey of the spectral type. This result is similar to that also reported by \citet[][0.7\%]{butler04} based on a survey of 150 M dwarfs with potential overlap with the \citet{endl06} sample. There have been no detections of so-called ``hot Jupiters'' (\textsl{a} $\sim$ 0.04 AU) around M dwarfs despite the stronger sensitivity of the Doppler detection method to these types of planets around low-mass stars. Around FGK-type stars, \citet{marcy05} cites a frequency of 1.2\% for hot Jupiters.

While the limits that can be currently placed on the frequency of Jupiter-mass planets around M dwarfs are not entirely inconsistent with those of FGK-type stars, there does seem to be a trend to fewer high-mass planet detections around M dwarfs. If further results support this observation, it would be consistent with the predictions of the core accretion planet formation model. \citet{laughlin04} and \citet{ida05} have shown that the formation probability of high-mass planets decreases with stellar mass with this model. Conversely, \citet{boss06} suggests that giant planets might actually form more efficiently around M dwarfs if the gravitational instability mechanism is considered. 

Further clouding the issue is the question of host star metallicity. As mentioned earlier, planets are more often detected around stars with high metallicities. In this Letter we have presented the results from our spectroscopic metallicity analysis of three M dwarfs that harbor extrasolar planets. We find that all three have sub-solar metallicities which is a departure from the observed trend in the FGK-type stars that harbor extrasolar planets. In contrast, \citet{bonfils05a} presented metallicity measurements for these M dwarfs based on a lower precision photometric relationship. Our derived [M/H] values are lower by 0.09, 0.08, and 0.34 dex for GJ 876, GJ 581, and GJ 436 respectively than those determined by \citet{bonfils05a}. In the case of the first two, the values are well within the overlapping errors for the two measurements (0.32 dex), while our measurement for GJ 436 is just outside this differential range. 

Taken together, the results from our analysis and that of \citet{bonfils05a} do appear to rule out super-solar metallicities for these M dwarf planet hosts. This result raises some interesting questions. Are the metallicities for these stars representative of the metallicities of the M dwarfs on planet search programs and might that explain the lower detection rates of planets for the M dwarfs? If that were the case, are the solar neighborhood M dwarfs in general metal deficient relative to the other spectral type? Or, what is causing the selection effect to lower metallicity M dwarfs for the planet search programs? 

We plan to follow up this preliminary result by applying the same spectroscopic metallicity analysis technique to the M dwarfs that have been surveyed for extrasolar planets as described by \citet{endl06}. The results from a larger sample should make it easier to disentangle the effects of stellar mass and metallicity on planet formation.

\acknowledgments
We thank Chris Sneden for his careful reading of the manuscript and excellent suggestions which improved this paper. GFB and JLB acknowledge support from NASA GO-06036, GO-06047, GO-06764, GO-08292, GO-08729, GO-08774, GO-09234, GO-09408, and GO-10773 from the Space Telescope Science Institute, which is operated by the Association of Universities for Research in Astronomy, Inc., under NASA contract NAS5-26555; and from JPL 1227563 ({\it SIM} MASSIF Key Project, Todd Henry, P.I.), administered by the Jet Propulsion Laboratory. ME is supported by the National Aeronautics and Space Administration under Grant NNG04G141G. The Hobby-Eberly Telescope is a joint project of the University of Texas at Austin, the Pennsylvania State University, Stanford University, Ludwig-Maximilians-Universit\"{a}t Muenchen, and Georg-August-Universit\"{a}t G\"{o}ttingen. The HET is named in honor of its principal benefactors, William P. Hobby and Robert E. Eberly.

\clearpage
\begin{deluxetable}{ccccccc}
\tabletypesize{\scriptsize}
\tablecolumns{7}
\tablewidth{0pc}
\tablecaption{Spectral types and derived stellar parameters for the M dwarf planet hosts.}
\tablehead{
 \colhead{Name} &
 \colhead{Spectral Type} &
 \colhead{\textsl{T}$_{eff}$\tablenotemark{a}} &
 \colhead{log \textsl{g}} &
 \colhead{[M/H]} &
 \colhead{$\xi$} &
 \colhead{$\eta$} \\
 \colhead{} &
 \colhead{} &
 \colhead{(K)} &
 \colhead{(cgs)} &
 \colhead{} &
 \colhead{(km s$^{-1}$)} &
 \colhead{(km s$^{-1}$)} 
}
\startdata
GJ 876  & M4   & 3478 & 4.89 & -0.12 & 0.77 & 0.64 \\
GJ 436  & M2.5 & 3498 & 4.80 & -0.32 & 1.02 & 0.00 \\
GJ 581  & M3   & 3480 & 4.92 & -0.33 & 0.91 & 1.35 \\
\enddata
\label{tab:table1}
\tablenotetext{a}{Adopted uncertainties are 48 K, 0.10 dex, 0.12 dex, 0.15 km s$^{-1}$, and 0.20 km s$^{-1}$ for the derived parameters {T}$_{eff}$, log \textsl{g}, [M/H], $\xi$, and $\eta$ respectively.}
\end{deluxetable}

\clearpage

\clearpage

\end{document}